\documentclass[mathleft
]{an}
\usepackage{graphicx}
\usepackage{times}
\usepackage{natbib}
\bibpunct{(}{)}{;}{a}{}{,}
\begin{document}

\Pagespan{1}{}
\Yearpublication{2016}%
\Yearsubmission{2016}%
\Month{}%
\Volume{}%
\Issue{}%

\title{Perspectives for observing hot massive  stars with {\it XMM-Newton} in the years 2017 -- 2027}

\author{G.\ Rauw\inst{1}\fnmsep\thanks{\email{rauw@astro.ulg.ac.be}\newline}}
\titlerunning{Hot stars in the next decade}
\authorrunning{G.\ Rauw}
\institute{Space sciences, Technologies and Astrophysics Research (STAR) Institute, Universit\'e de Li\`ege, All\'ee du 6 Ao\^ut, 19c, B\^at B5c, 4000 Li\`ege, Belgium}

\received{}
\accepted{}
\publonline{}


\abstract{{\it XMM-Newton} has deeply changed our picture of X-ray emission of hot, massive stars. High-resolution X-ray spectroscopy as well as monitoring of these objects helped us gain a deeper insight into the physics of single massive stars with or without magnetic fields, as well as of massive binary systems, where the stellar winds of both stars interact. These observations also revealed a number of previously unexpected features that challenge our understanding of the dynamics of the stellar winds of massive stars. I briefly summarize the results obtained over the past 15 years and highlight the perspectives for the next decade. It is anticipated that coordinated (X-ray and optical or UV) monitoring and time-critical observations of either single or binary massive stars will become the most important topics in this field over the coming years. Synergies with existing or forthcoming X-ray observatories ({\it NuStar}, {\it Swift}, {\it eROSITA}) will also play a major role and will further enhance the importance of {\it XMM-Newton} in our quest for understanding the physics of hot, massive stars.}

\maketitle

\section{Introduction}
Hot stars with initial masses of more than 10\,M$_{\odot}$ populate the upper left part of the Hertzsprung-Russell diagram, corresponding to spectral types from early-B to O. These objects evolve over short lifetimes (typically of order a few to ten Myr) and appear as Wolf-Rayet (WR) stars during the advanced stages of their evolution. Despite their rather short lifetimes, these stars have a tremendous impact on the interstellar medium (ISM) and the ecology of their host galaxies as a whole. Owing to their high photospheric temperatures and huge luminosities, they release copious amounts of UV photons, thereby dominating the ionization of the interstellar medium \citep[e.g.][]{Reynolds}. These intense radiation fields further drive powerful outflows in the form of stellar winds \citep[e.g.][]{Puls}, associating large terminal velocities (1000 -- 3000\,km\,s$^{-1}$) and huge mass-loss rates ($10^{-7}$ -- $10^{-4}$\,M$_{\odot}$\,yr$^{-1}$). Mass-loss via these winds plays a significant role in the evolution of massive stars \citep[e.g.][]{Hirschi}. Furthermore, these stellar winds inject huge amounts of mechanical energy into the ISM and, especially during the WR stage, contribute to the chemical enrichment of the interstellar gas \citep[e.g.][]{Langer}. In massive binary systems, the collision between the stellar winds leads to hydrodynamical shocks that can act as acceleration sites for relativistic particles, as evidenced by the detection of synchrotron radio emission for a subset of massive binaries \citep[e.g][]{Benaglia}. Therefore, studying the physics of stellar winds is not only important for the understanding of massive stars, but also for quantifying their role in feedback processes at the scales of star-forming regions or galaxies.\\ 

X-ray emission from massive early-type stars has been known for almost four decades now \citep{Harnden,Seward}. From early studies with the {\it EINSTEIN} and {\it ROSAT} observatories, it became clear that the X-ray luminosity of O-type stars scales with their bolometric luminosity according to $L_{\rm X}/L_{\rm bol} \sim 10^{-7}$ \citep[e.g.][]{LW,Berghoefer} and that the X-ray emission is usually rather soft ($kT < 1$\,keV) with little circumstellar absorption ($N_{\rm H}^{\rm wind} \simeq $ a few $10^{21}$\,cm$^{-2}$), unlike what one would expect if the X-rays were coming from a hot, magnetically driven, corona at the base of the dense stellar wind of these stars \citep{LW}. These results led to the paradigm that the X-ray emission of massive stars must arise inside their stellar winds as a result of intrinsic instabilities of these winds that develop into hydrodynamic shocks via the so-called line deshadowing instability \citep[e.g.][]{Owocki88,Feldmeier,RO}. The situation for WR stars was found to be less clear: \citet{Pollock} showed that X-ray bright WR stars are binaries, whilst single WR stars are usually faint, with nitrogen-rich (WN) Wolf-Rayet stars being brighter on average than carbon-rich (WC) Wolf-Rayet stars. Contrary to what happens for O-type stars, no clear dependence between X-ray and bolometric luminosities exists for WR stars \citep{Wessolowski,Ignace00}.

In parallel, it was found that some massive binaries are significantly brighter than expected from their bolometric luminosity and exhibit variability along their orbit \citep[e.g.][]{Pollock,CG}. These results revived the idea, initially formulated by \citet{PU} and \citet{Chere}, that an excess X-ray emission arises from the hot shocked plasma in the wind-wind interaction zone between the components of massive binary systems \citep[e.g.][]{SBP}. Since massive stars are moderately bright X-ray sources, that are mostly located at distances of a few kpc, substantial progress in the study of these objects was expected from the gain in sensitivity and spectral resolution that was achieved with the launch of the {\it XMM-Newton} and {\it Chandra} observatories. Indeed, many new discoveries have been made over the past sixteen years and some of them are briefly highlighted in Sect.\,\ref{review}\footnote{Unfortunately, within the present paper, it is impossible to review all studies in detail. For more extensive recent reviews on X-ray emission of massive stars, I refer the reader to \citet{GN}, and the recent special issue of Advances in Space Research (volume 58, September 2016).}. Section\,\ref{future} addresses some topics that are likely to play a key role in future studies of massive stars with {\it XMM-Newton}. Finally, Sect.\,\ref{conclusions} summarizes the conclusions.

\section{What we have learned so far\label{review}}
\subsection{Single, non-magnetic, massive stars}
A handful of presumably single non-magnetic massive stars are sufficiently bright in X-rays to study them with the current generation of high-resolution X-ray spectrographs. This allowed to resolve the profiles of individual emission lines and to confront them with theoretical predictions. On theoretical grounds, it is expected that the profiles should be skewed due to the absorption by the stellar wind affecting more heavily the red side of the profile \citep[e.g.][]{Macfarlane,Feldmeier03,OC}. This opens up the possibility to estimate the optical depth of the winds, and hence to infer the mass-loss rates \citep[e.g.][]{Cohen}. Yet, the complex interplay between contributions from plasma at different temperatures and with different locations inside the wind, on the one hand, and absorption by a clumpy cool wind, on the other hand, leads to quite complex situations \citep[e.g.][]{Oskinova16}. Therefore, it is necessary to analyse the high-resolution X-ray spectra along with high-resolution optical and UV spectra to achieve a consistent picture. Indeed, multi-wavelength data allow to lift the degeneracy between the mass-loss rate and the porosity of the stellar wind that affects X-ray emission line profiles. Promising results have been obtained through global fits of the RGS spectra of $\zeta$\,Pup \citep[O4\,Ief,][]{Herve} and $\lambda$\,Cep \citep[O6\,Ief,][]{lamCep} combined with the results of optical/UV spectroscopy. Whilst it was found that the mass-loss rates inferred in the joint optical/UV analyses allow to reproduce the RGS spectra, the X-ray data indicated a location of the X-ray plasma much closer to the photosphere than anticipated \citep[down to 1.2\,$R_*$, e.g.][]{lamCep}. This latter result shows that X-rays are generated already in the innermost regions of the wind, where the outflow is still accelerating according to the standard theory \citep{CAK}. These findings support results from recent hydrodynamical simulations by \citet{SO} that predict the onset of the line deshadowing instability already near 1.1\,$R_*$.

\begin{figure*}
\begin{minipage}{13cm}
\includegraphics[width=13cm]{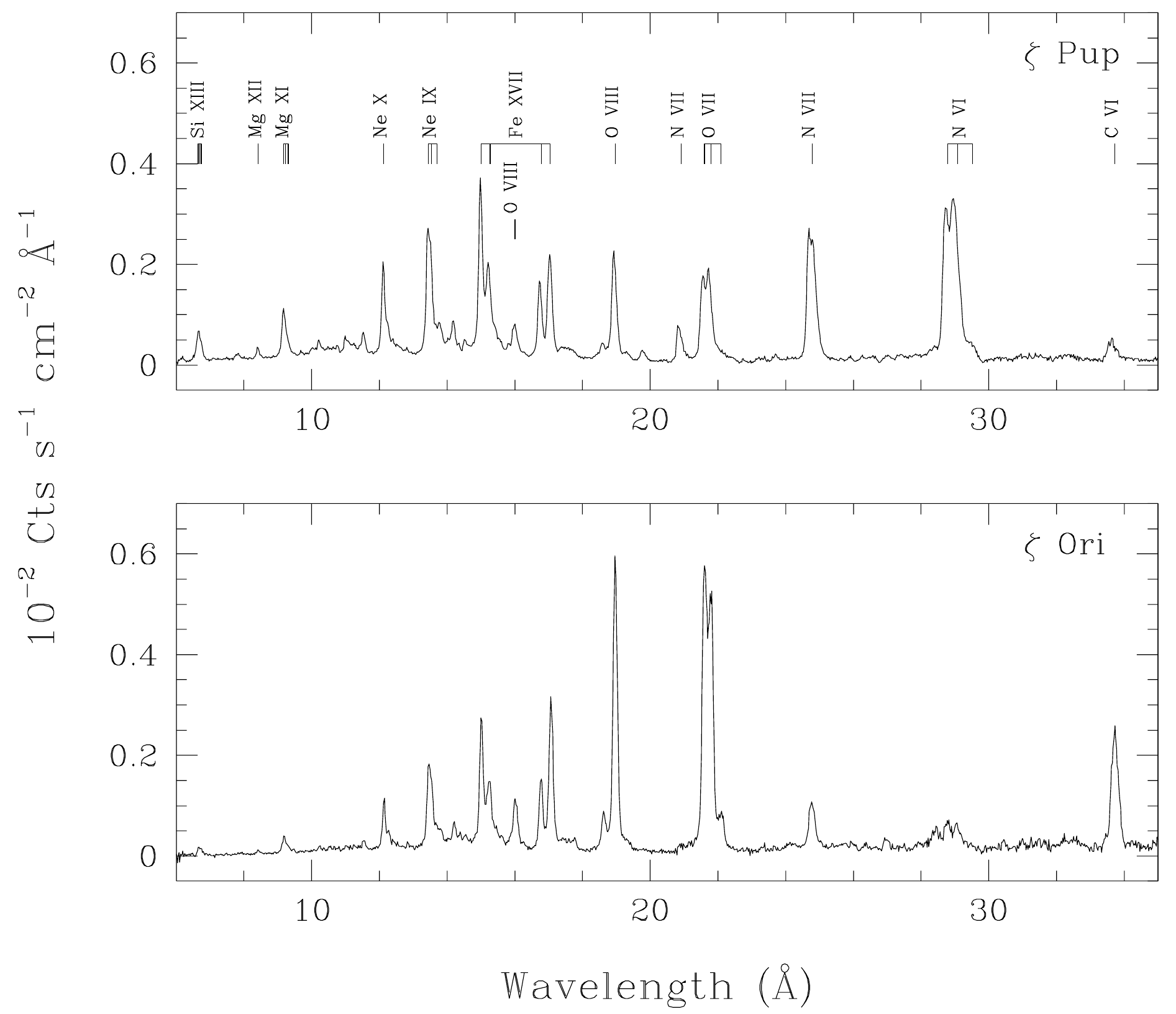}
\end{minipage}
\begin{minipage}{4cm}
\caption{RGS spectra of the two most-extensively observed O-type stars, $\zeta$\,Pup (top) and $\zeta$\,Ori (bottom). Note the very different relative strengths of the CNO lines in the spectra of these stars.}
\label{label1}
\end{minipage}
\end{figure*}

These joint optical/UV and X-ray analyses also provide insight into the evolutionary stage of the stars. Figure\,\ref{label1} illustrates the clear differences in the relative strengths of the C, N and O lines between the two supergiants $\zeta$\,Pup (O4\,Ief) and $\zeta$\,Ori (O9.7\,Ib). The former star clearly presents an enhancement of its nitrogen abundance along with a depletion of oxygen and carbon that reflect the presence of CNO-processed material at the stellar surface \citep{Herve,Kahn}. 

As single Wolf-Rayet stars are X-ray fainter than O-stars, the WN4 star WR\,6 is currently the only WR star that could be observed at high-resolution in X-rays with the RGS aboard {\it XMM-Newton} and the HETG aboard {\it Chandra} \citep{OskinovaWR6,Huenemoerder}. The X-ray lines were resolved and the HETG data revealed broad, blue-shifted X-ray lines. This morphology indicates line formation in the outer regions (tens to hundreds of $R_*$) of a uniformly expanding wind, just above the radius where the radial optical depth of the wind for X-rays becomes unity \citep{Huenemoerder}. Such a large formation radius strongly contrasts with the case of OB stars, suggesting that both phenomena arise through different mechanisms. However, \citet{Gayley} presented a first attempt to unify the X-ray generation processes in OB and WR winds. He suggests that the detection of X-ray emission of WR stars and hence the location of their observable X-ray emitting plasma depend on the competition between photoelectric absorption by the dense wind and the capability of the wind to advect turbulent gas to outer regions.\\  

The X-ray variability of single massive stars was investigated mostly using EPIC data from the longest available {\it XMM-Newton} exposures. The excellent quality of the {\it XMM-Newton} time series of $\zeta$\,Pup allowed to study the variability even on short timescales. The absence of significant short-term variability yields stringent constraints on the number of small-scale structures: \citet{zetaPup} showed that the wind of $\zeta$\,Pup is highly fragmented, containing at any time more than $10^5$ independent clumps. Contrasting with the lack of strong short-term variability, there is mounting evidence for variability on timescales of the rotation period. Strong hints for modulations of the X-ray emission at the 10 -- 15\% level have been found for the O-type stars $\zeta$\,Pup \citep{zetaPup}, $\xi$\,Per \citep[O7.5\,III(n)((f)),][]{Massa} and $\lambda$\,Cep \citep{lamCep}. These variations could arise from large-scale spiral structures in the wind co-rotating with the star.

In the same context, \citet{Ignace} discuss $4 \times 100$\,ks EPIC data of WR\,6. This star presents a well established 3.766\,day cycle at other wavelengths, but the phased X-ray light curve fails to confirm obvious cycles. Yet WR\,6 produces X-ray variability at the 10 - 20\% level. A co-rotating interaction region could explain the observed variability, but the lack of a strictly periodic behaviour would require a perturbation of the structure by a kind of wave propagating along the structure. 
\subsection{Magnetic OB stars}
Using the current generation of spectropolarimetric devices, it has been shown that about 6 - 8\% of massive stars host essentially dipolar magnetic fields \citep[e.g.][]{wad13,fos15}. The interplay between the stellar wind and this magnetic field produces a so-called magnetically confined wind with strong shocks leading to a hot X-ray emitting plasma confined near the magnetic equator \citep[and references therein]{Babel,udDoula}. \citet{YNBfield} performed a survey of the X-ray properties of about two thirds of the known magnetic OB stars, spanning a wide range in magnetic parameter space. These authors showed that the semi-analytical model of \citet{udDoula} predicts the observed X-ray luminosities rather well. \cite{YNBfield} further showed that the X-ray emission is not necessarily very hard and that there exists no correlation between the spectral hardness and the magnetic or stellar parameters. A comprehensive review of our current understanding of the X-ray properties of magnetic OB stars can be found in \citet{AsifYael}. 
As the axis of the magnetic dipole is often inclined with respect to the rotation axis, one frequently observes a phase-locked rotational modulation of the optical (H$\alpha$), UV and X-ray emission. The observed variations depend on the magnetospheric geometry and the orientation of the star with respect to the observer. The most likely reason for the X-ray variability is occultation of parts of the magnetosphere by the stellar body though the details need to be understood \citep[and references therein]{AsifYael}. Monitoring the X-ray emission is demanding because of the time constraints and requires rather large amounts of observing time. However, it is clearly an excellent means to probe the magnetosphere of massive stars and obtain unique constraints on the magneto-hydrodynamic properties of its plasma, especially when combined with coordinated optical and/or UV spectroscopy \citep[e.g.][]{CPD}.

A most interesting result was the detection of X-ray pulsations in the magnetic B0.5\,IV star $\xi^1$\,CMa \citep{xiCMa,YNxi1}. This star is a known non-radial pulsator of $\beta$\,Cep-type with a 4.9\,hrs pulsation period. The broadband X-ray flux of $\xi^1$\,CMa was found to exhibit a 10\% modulation on the same period and in phase with the non-radial pulsations. Furthermore, a possible modulation of the N\,{\sc vii} Ly$\alpha$ line was also found \citep{xiCMa} though this latter result needs confirmation. A connection between the pulsations, at the level of the photosphere, and the X-ray emission, generated in the stellar wind, suggests that the pulsations trigger part of the wind instabilities. In a subsequent study, \citet{Oskinova15} investigated archival {\it XMM-Newton} and {\it Chandra} data of three other $\beta$~Cep stars ($\kappa$~Sco, $\beta$~Cru, $\alpha$~Vir). Whilst weak indications of X-ray variability were found in $\beta$~Cru, no statistically significant evidence of pulsations was obtained. It has to be stressed though that these three objects have significantly lower count rates than $\xi^1$~CMa, making it more difficult to achieve a good detection sensitivity and rendering the detection of a 10\% modulation (as seen in $\xi^1$\,CMa) impossible.

\begin{figure}
\includegraphics[width=8.2cm]{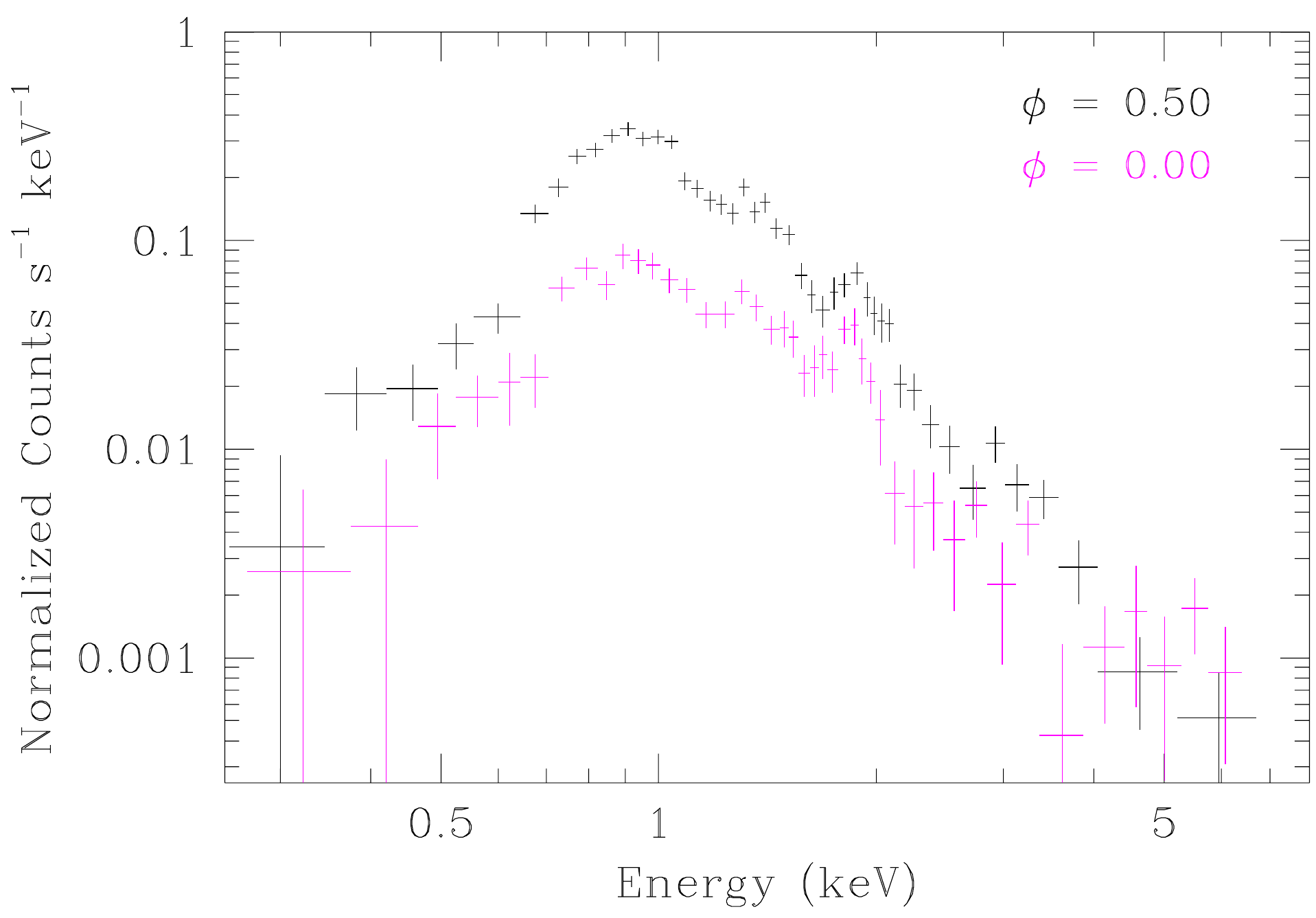}
\caption{EPIC-pn spectra of the HDE\,228766 binary ($P_{\rm orb} = 10.7$\,days, $e = 0$) at the conjunction phases with the Of$^+$/WN8ha star in front of the O7\,III-I companion ($\phi = 0.00$) and the opposite configuration ($\phi = 0.50$). The Of$^+$/WN8ha star is found to have a wind momentum that exceeds that of its companion by at least a factor 5  \citep{Rauw14}. The effect of the changing line-of-sight between the two conjunction phases is clearly seen through the changing flux level at energies below 2\,keV.}
\label{fig3}
\end{figure}
\subsection{Colliding wind binaries}
Massive binaries host wind interaction regions that are known to produce additional X-ray emission. The observable emission varies with orbital phase due to changing line-of-sight optical depth and/or changing orbital separation in eccentric systems (\citealt{SBP}; \citealt{PP}; see \citealt{GRYN}, for a recent review). At the end of the 20th century, our knowledge of colliding wind binaries was very incomplete. The current generation of X-ray observatories has deeply changed our observational picture of these systems. 

Studying large samples of massive stars containing O-star binaries, it was found that not all massive binaries are X-ray overluminous compared to the canonical $L_{\rm X}/L_{\rm bol}$ relation \citep[e.g.][]{Sana,YN,CygOB2}. Hence, the question that comes up is why some systems are X-ray bright, whilst others do not show any clear signs of X-rays coming from the wind interaction region. This issue is currently still open and requires further multi-wavelength studies to identify the parameter(s) and phenomena that rule the level of the X-ray emission produced in the wind interaction zone.  

Meanwhile, X-ray light curves of colliding wind binaries provide an efficient means to study the hydrodynamics of stellar winds. For systems with circular orbits and hosting stars with very different wind strengths, comparing the X-ray spectra at opposite conjunction phases allows to establish the variation of the line-of-sight absorption towards the wind interaction region and hence to derive constraints on the wind properties \citep[e.g.\ the O7\,III-I + Of$^+$/WN8ha system HDE\,228766,][see Fig.\,\ref{fig3}]{Rauw14}. 

{\it XMM-Newton} and {\it Swift} have allowed to gather phase-resolved X-ray spectroscopy of a sample of colliding wind binaries that revealed more complex variations than expected a priori \citep[see][and references therein]{GRYN}. \citet{Lomax} studied the light curve of V444\,Cyg (WN5 + O6, $e = 0$, $P_{\rm orb} = 4.2$\,d) and interpreted the asymmetric behaviour of the X-ray light curve around the eclipses as evidence of the Coriolis distortion of the wind-wind collision. The unexpectedly large opening angle of the X-ray emitting region and its location provide evidence of radiative inhibition and/or braking \citep{SP,GOC} occurring within this system \citep{Lomax}. In relatively wide eccentric binaries, the wind-wind interactions are expected to be in the adiabatic regime and consequently the X-ray emission of the wind interaction zone is expected to vary as $1/d$ where $d$ is the orbital separation \citep{SBP}. Such a situation was indeed observed by \citet{CygOB2n9} in the 2.35\,yr eccentric ($e = 0.71$) O5-5.5\,I + O3-4\,III binary Cyg\,OB2 \#9. However, in other systems, such as 9~Sgr \citep[O3.5 V((f$^*$)) + O5-5.5 V((f)), $e = 0.71$, $P_{\rm orb} = 9.1$\,yr;][]{9Sgr} and WR\,25 \citep[WN6h + O4\,f, $e = 0.50$, $P_{\rm orb} = 208$\,days;][]{Pandey}, deviations from this simple behaviour have been observed. Even stronger deviations from the $1/d$ scaling are seen in shorter period eccentric systems, where the wind interaction zone can switch from mostly adiabatic near apastron to highly radiative near periastron. Theoretical models of \citet{PP} indeed predict the X-ray flux at a given orbital phase to depend on the history of the plasma heated at earlier orbital phases, resulting in a hysteresis-like variation of the flux as a function of orbital separation. Observations with {\it XMM-Newton} and {\it Swift} have revealed this kind of behaviour \citep{Cazorla,GRYN}. A very interesting  example of a hysteresis-like loop is found for the 31.7\,day eccentric ($e = 0.69$) binary WR\,21a (WN5h + O3\,V). Using {\it Swift} and {\it XMM-Newton} observations, \citet{GN16} showed that the X-ray emission of this system follows a $1/d$ trend between orbital phases 0.2 and 0.9. After $\phi = 0.9$,  the emission steeply declines both in the soft and hard bands. This is likely due to a strong absorption by the Wolf-Rayet wind, combined with a collapse of the colliding wind region near periastron. The emission then slowly recovers as the stars move away from periastron, closing the hysteresis cycle.   
 
\section{The next decade \label{future}}
Whilst important progress has been made in our understanding of the X-ray properties of hot massive stars, there are many remaining open questions that can be addressed with {\it XMM-Newton} in the coming years. There is still a handful of massive OB stars that are within reach of the RGS spectrograph although exposure times near 300\,ks are needed to achieve a good signal-to-noise ratio. Enlarging the currently limited sample of single O-type stars with decent quality high-resolution X-ray spectra is important to constrain the location of the X-ray emitting plasma inside the winds and to check whether the generation of X-rays close to the stellar photosphere is a general property of massive stars. 

Owing to its instrumental stability and the fact that data are collected simultaneously with the three EPIC cameras and the two RGS spectrographs, {\it XMM-Newton} is a unique tool for studies in the time domain. This offers excellent opportunities for in-depth studies of stellar winds, notably 
\begin{itemize} 
\item[$\bullet$] via the study of structures in the winds of X-ray bright O or WR stars in coordination with ground-based (optical) or space-borne (UV) spectroscopy or photometry. As outlined above, this might be one of the best ways to investigate the connection between the photosphere and the wind. 
\item[$\bullet$] through monitoring of critical orbital phases of selected colliding wind binary systems to probe the physics of the shocks. The existence of hysteresis-like cycles in eccentric systems implies that observing one half of the orbital cycle is not sufficient, as data from symmetric orbital phases might correspond to different physical stages of the shock-heated plasma.
\item[$\bullet$] via monitoring of the rotational modulation of the X-ray emission from hot stars with magnetically confined winds. This allows to map the hot plasma in their magnetospheres. 
\end{itemize}

Synergies with other X-ray facilities offer further possibilities. For instance, {\it Swift} can be used to monitor colliding wind systems in support to more detailed spectroscopy with {\it XMM-Newton} at critical orbital phases \citep[see e.g.\ the studies of][]{Lomax,Pandey,GN16}. 

\begin{figure}
\includegraphics[width=8.2cm]{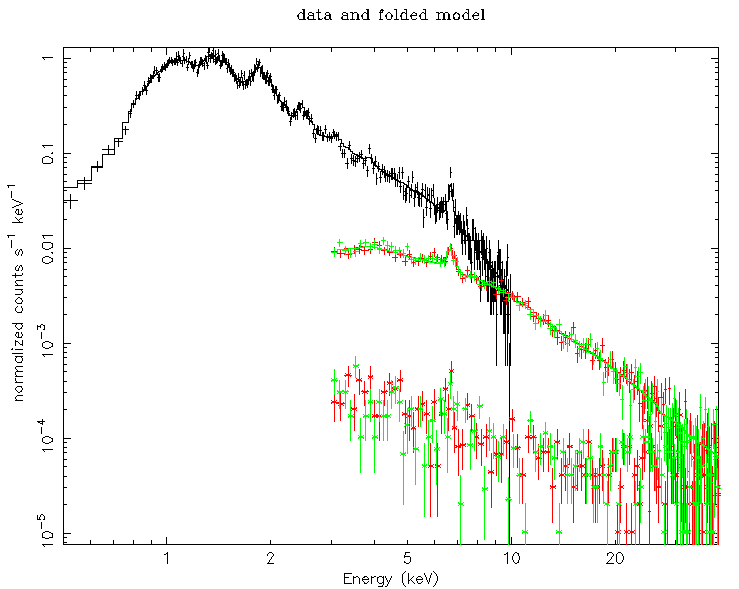}
\caption{Simulated 100\,ks {\it NuSTAR} spectra (red and green symbols, one spectrum for each of the two focal plane modules) and 20\,ks {\it XMM-Newton} EPIC-pn spectrum (black symbols) of the colliding wind system Cyg\,OB2 \#8a (O6\,If + O5.5\,III(f), $P_{\rm orb} = 21.9$\,days, $e = 0.24$). The model uses the thermal spectrum as observed with {\it XMM-Newton} \citep{Cazorla} and assumes a power-law with $\Gamma = 1.5$ and with a 20--60\,keV flux of $6.1 \times 10^{-12}$\,erg\,cm$^{-2}$\,s$^{-1}$, corresponding to the upper limit inferred by \citet{FDP} from {\it INTEGRAL} data. The simulated {\it NuSTAR} background is also shown for comparison.}
\label{fig2}
\end{figure}

{\it NuSTAR} allows to study the hard tails of $\gamma$\,Cas analogs\footnote{For a detailed review of these intriguing hard X-ray emitting Be/Oe stars, see e.g.\ \citet{Smith}.} and of colliding wind systems. Hard non-thermal power-law components are expected to form through inverse Compton scattering in colliding wind binary systems displaying synchrotron radio emission. Whilst {\it NuSTAR} is the ideal tool to detect such a hard tail, coordinated observations with {\it XMM-Newton} are needed to fully constrain the properties of the thermal plasma from spectra at lower energies and to assess the contribution of this thermal emission to the hard X-ray tail (see Fig.\,\ref{fig2}). For instance, \citet{Hamaguchi} present joint {\it XMM-Newton} and {\it NuSTAR} observations of the colliding wind system $\eta$~Car near periastron passage. The {\it NuSTAR} data revealed a hard tail up to 50\,keV consistent with thermal bremsstrahlung emission arising in a $\sim 6$\,keV plasma. No power-law component was found in the very hard band (unlike what {\it INTEGRAL} and {\it Suzaku} had detected before). 

Finally, {\it eROSITA} will provide a sensitive survey of a large sample of massive stars in our Galaxy, thereby identifying interesting targets (e.g.\ full samples of $\gamma$\,Cas analogs) for deep follow-up observations with {\it XMM-Newton}. 
\section{Conclusions \label{conclusions}}
Over the last two decades, the high spectral resolution and excellent monitoring capabilities of {\it XMM-Newton} tremendously improved our understanding of the physics of hot massive stars and their X-ray emission. X-ray spectroscopy and variability studies appear nowadays as one of the most promising avenues to study the connection between the photosphere and the wind. 
Still, many open issues related to the physics of shocks in stellar winds remain. These questions can be addressed in the coming years with {\it XMM-Newton}.
Important aspects include:
\begin{itemize}
\item[$\bullet$] studies in the time domain (colliding winds, variability of single stars,...) to investigate the properties of stellar winds of single and binary massive stars
\item[$\bullet$] synergies with other X-ray facilities and coordinated observations with ground-based telescopes (including medium-sized private telescopes that offer a better flexibility than very large telescopes) or space-borne observatories (e.g.\ {\it HST} for UV spectroscopy, but also much smaller facilities such as {\it MOST} or {\it BRITE} for high-precision photometry).
\end{itemize}
Some of these topics do require substantial amounts of observing time (to cover the relevant time scales and/or to achieve the requested signal-to-noise ratios). However, they are clearly worth the effort as they will provide deep insight into the physics of stellar winds of massive stars and thereby further our understanding of feedback processes at the scale of stellar clusters and galaxies.

\acknowledgements
I would like to thank my colleague Ya\"el Naz\'e for discussion, and express my gratitude to the {\it XMM-Newton}-SOC for their support, notably in scheduling time-constrained observations. I acknowledge financial support from Li\`ege University, through an ARC grant for Concerted Research Actions, financed by the French Community of Belgium (Wallonia-Brussels Federation), from the Fonds de la Recherche Scientifique (FRS/FNRS), as well as through an XMM PRODEX contract (Belspo).

\end{document}